\documentstyle[prb,aps,epsfig,draft,floats,amsfonts]{revtex}

\begin{document}

\twocolumn \psfull \draft

\wideabs{
\title{Structure of quantum disordered wave functions: weak
localization, far tails, and mesoscopic transport}
\author{Branislav K. Nikoli\' c$^1,*$ and Viktor Z. Cerovski$^2$}
\address{$^1$Department of Physics, Georgetown University,
Washington, DC 20057-0995 \\ $^2$Department of Physics, Virginia
Commonwealth University, Richmond, VA 23284}

\maketitle

\begin{abstract}
We report on the comprehensive numerical study of the fluctuation
and correlation properties of wave functions in three-dimensional
mesoscopic diffusive conductors. Several large sets of nanoscale
samples with finite metallic conductance, modeled by an Anderson
model with different strengths of diagonal box disorder, have
been generated in order to investigate both small and large
deviations (as well as the connection between them) of the
distribution function of eigenstate amplitudes from the universal
prediction of random matrix theory. We find that small, weak
localization-type, deviations contain both diffusive
contributions (determined by the bulk and boundary conditions
dependent terms) and ballistic ones which are generated by
electron dynamics below the length scale set by the mean free
path $\ell$. By relating the extracted parameters of the
functional form of nonperturbative deviations (``far tails'') to
the exactly calculated transport properties of mesoscopic
conductors, we compare our findings based on the full solution of
the Schr\"odinger equation to different approximative analytical 
treatments. We find that statistics in the far tail can be
explained by the exp-log-cube asymptotics (convincingly refuting
the log-normal alternative), but with parameters whose dependence
on $\ell$ is linear and, therefore, expected to be dominated by
ballistic effects. It is demonstrated that both small deviations
and far tails depend explicitly on the sample size---the
remaining puzzle then is the evolution of the far tail parameters
with the size of the conductor since short-scale physics is
supposedly insensitive to the sample boundaries.
\end{abstract}

\pacs{PACS numbers: 73.23.-b, 72.15.Rn,  05.45.Mt, 05.40.-a}}

\narrowtext

\section{Introduction} \label{sec:intro}

Quantum coherence, its nonlocal features, and randomness of
microscopic details can cause large fluctuations of physical
quantities in disordered mesoscopic systems. The paradigmatic
case is that of conductance fluctuations which has given impetus
for the whole field of mesoscopic physics~\cite{meso,mqp} by
pointing out at unexpected features of such
fluctuations.~\cite{ucf} Contrary to the intuition developed from
thermal fluctuations, and their self-averaging properties in the
statistical physics of macroscopic systems, the  average value
and variance are not enough to characterize the broad
distribution functions of various mesoscopic
quantities,~\cite{shapiro}  even well into the metallic regime $g
\gg 1$ ($g=G/G_Q$ being the dimensionless zero-temperature
conductance, in units of conductance quantum $G_Q=2e^2/h$). The
fluctuations increase, broadening the distributions, as disorder
is increased eventually driving a system through the
localization-delocalization (LD) transition~\cite{anderson} at $g
\sim 1$. Thus, the mesoscopic program was born where full
distribution functions of relevant quantities in open (e.g.,
conductance, local density of state, current relaxation times,
etc.) or closed (e.g., eigenfunction amplitudes, polarizability,
level curvatures, etc.) samples are to be
studied.~\cite{meso,janssen} Especially interesting are the large
deviations of their asymptotic tails from the ubiquitous Gaussian
distributions (which are expected only in the limit $g
\rightarrow \infty$).

Recently, the study of fluctuations and correlations of
eigenfunction amplitudes has been initiated.~\cite{mirlin} The
quantum coherence induces long-range spatial correlations (due to
massless modes, like diffusons and Cooperons) in the local density
of states and eigenfunction amplitudes, which in turn lead to
strong mesoscopic fluctuations of global quantities like
conductance. Small deviations of eigenstate statistics from the
universal predictions (applicable in  the limit $g \rightarrow
\infty$) of random matrix theory (RMT) are well understood
through perturbative corrections $\sim {\mathcal O}(g^{-1})$ of
the weak localization (WL) type,~\cite{fyodorov} but the physical
origin of large deviations in the far asymptotic tail of the
distribution function is much more controversial.~\cite{mirlinsusy,efetov,smolyarenko}
Not only that there are different analytical predictions  for the
far tail asymptotics (which in fact do not explain all details of the
tails found in numerical simulations~\cite{uskiprb,bkn}), but
there is also an issue~\cite{smolyarenko} of the relevant physics
which is responsible for large wave function amplitudes (quantum
vs semiclassical) and a closely related question on the limitations of
usually employed field-theoretical approaches~\cite{efetovbook} to
study the disordered electron problem. Also, the parameters of the
WL correction to the RMT framework cannot be explained
(e.g., in the Anderson model~\cite{bkn}) solely by the standard universal
(independent of the details of disorder) quantities extracted from the
semiclassical diffusive dynamics.~\cite{mirlin} Instead, careful examination
of ballistic effects, generated by the properties of particle dynamics on
the length scale below the mean free path $\ell$, is required.~\cite{mirlinbal}
Moreover, it is possible that some types of disorder could generate appreciable
higher order terms, (characterizing non-Gaussian features of random
potential~\cite{smoly_private}) in this perturbative expansion in
$1/g$, and thereby change the functional form of the perturbative
correction as well. Thus, a detailed study of deviations from the
RMT statistics in the paradigmatic case of a quantum particle in a
random potential offers a possibility to unravel underlying
correlations in a controlled fashion, which paves the way for
understanding plethora of diverse problems (including those
outside of physics~\cite{bouchaud}) where matrices containing random
elements and their eigenstates are encountered.

In the course of exploration of mesoscopic fluctuations, the
so-called prelocalized states have been unearthed as the
microscopic origin of asymptotic tails of various quantum
distribution functions of thermodynamic and kinetic
quantities.~\cite{efetov,lerner,muz,basu} While typical wave
function is spread uniformly throughout a metallic sample of
volume $L^d$ with average amplitude $L^{-d/2}$ (up to inevitable
Gaussian fluctuations), the prelocalized state in 3D exhibits
much larger local amplitude splashes (on the top of the
homogeneous background $|\Psi({\bf r})|^2 \sim L^{-d}$) at some
points ${\bf r}$ within the sample.~\cite{mirlin,bkn} To obtain
the far tail of such distribution ``experimentally'' (e.g., in
microwave cavities of Ref.~\onlinecite{sridhar} or by numerical
simulations~\cite{bkn}) in ``realistic'' metallic systems, one
has to search for extremely rare disorder configurations where
quantum interference effects are able to generate highly unusual
eigenfunctions.

What is the relevance of prelocalized states for transport
experiments? Most of phenomena in good metallic disordered
conductors are semiclassical in nature. This means that
disorder-averaged properties, like conductance measured in
experiments or calculated in (quantum transport)
theory,~\cite{allen} are determined by the usual extended states
of uniform amplitude, formed in the typical fluctuations of the
random potential. However, recent experiments on quantum dots
(nanofabricated samples with well-resolved electron energy
levels) show that some transport properties, like fluctuations of
the tunneling conductance, can depend sensitively on the local
features of wave functions which couple the dot to external
leads.~\cite{dots} Also, to understand the excitation and
addition spectra of quantum dots one has to deal with the
statistics of Coulomb interaction matrix elements, which are
influenced by the eigenfunction amplitude
fluctuations.~\cite{ogam} By exploiting the correspondence
between the Schr\" odinger and Maxwell equations in microwave
cavities, it has become possible to probe directly the
microscopic structure of wave functions in quantum disordered or
quantum chaotic systems.~\cite{sridhar}

Here we undertake a comprehensive search, through numerical
simulations, for special disorder configurations in order to
investigate functional dependence of the parameters determining
eigenstate statistics on the disorder strength or sample size.
This is not just a `brute force' study culminating in a fitting
procedure of the observed distribution functions, but more
importantly, an attempt to quantify those effects which can lead to
substantial deviations from the RMT, departing even from the
standard semiclassical corrections to it. Namely, our results stem from the exact
solutions of the Schr\" odinger equation for a particle in a
random potential, and therefore provide a reference point for the
analytical approaches which usually integrate out some degrees of
freedom by focusing on the ``low energy'' sector of a full
theory.~\cite{mirlin} For this purpose, we also compute exactly the
transport properties of our finite-size samples, and relate them
to the parameters extracted from the fits of analytical formulas
to the perturbative and far tail interval of the eigenfunction
amplitudes. Mesoscopic physics intrinsically deals with
finite-size phase-coherent samples, and has led to efficient use
of different transport formalisms. Thus, we exploit the fact that
transport properties of a specific sample (simulated here as
nanoscale single band conductors) can be measured exactly on a
computer. Although our focus is primarily on the peculiar states
exhibiting the largest amplitudes (which generate far tails of the
statistics of eigenfunction amplitudes, as well as other mesoscopic
quantities), it becomes necessary to investigate thoroughly
the region of small eigenfunction amplitudes because of the
possibility that the same semiclassical quantities (like classical
diffusion propagator,~\cite{fmmodes} which we evaluate explicitly for the samples
with specific boundary conditions) might govern both
portions of the distribution function.~\cite{mirlin}

We have investigated five different ensembles~\cite{ensemble} of
mesoscopic samples, each containing 30000 weakly disordered
three-dimensional (3D) metallic conductors. Finite-size samples
are modeled by a tight-binding Hamiltonian (TBH)
\begin{equation}\label{eq:tbh}
  \hat{H}=\sum_{\bf m} \varepsilon_{\bf m}  |{\bf m} \rangle \langle {\bf m}| +
   \sum_{\langle {\bf m},{\bf n} \rangle} t_{{\bf m} {\bf n}}|{\bf m} \rangle \langle {\bf
  n}|,
\end{equation}
with nearest-neighbor hopping $t_{{\bf m} {\bf n}}=1$ (unit of
energy) between $s$-orbitals $\langle {\bf r} | {\bf m} \rangle =
\psi({\bf r}-{\bf m})$ located on sites ${\bf m}$ of a simple
cubic lattice of size $L=12a$ to $L=20a$ ($a$ is the lattice spacing).
Periodic boundary conditions are chosen in all directions. The
disorder is simulated by taking potential energy
$\varepsilon_{\bf m}$ to be a uniformly distributed random
variable, $-W/2 < \varepsilon_{\bf m} < W/2$, which is the standard
(Anderson) model in the localization theory.~\cite{anderson} The
impurity configurations vary from sample to sample for a given disorder
strength $W\in \{3.5,4.0,4.5,5.0,5.5\}$ which is chosen to ensure that
ensemble-averaged transport quantities characterize metallic
($g \gg 1$) diffusive ($\ell \ll L$) transport regime at half filling
(i.e., at Fermi energy $E_F=0$). For a weak $k_F \ell \gg 1$
($k_F$ is the Fermi wave vector) disorder, disorder-averaged transport
properties are semiclassical, i.e., well-described by the
Bloch-Boltzmann formalism. Going beyond $W \simeq 6$ (but below
critical $W_c \simeq 16.5$ for the localization of the whole
band) would still give metallic conductance $g \gg 1$ (for
large enough lattice), but the semiclassical concepts appearing
in analytical predictions for the far tail,~\cite{mirlin} like
$\ell$, loose their meaning.~\cite{allen} The disorder strengths
below $W=3$ are excluded only because of requiring too large lattices
to avoid quasiballistic transport.
\begin{figure}
\centerline{ \psfig{file=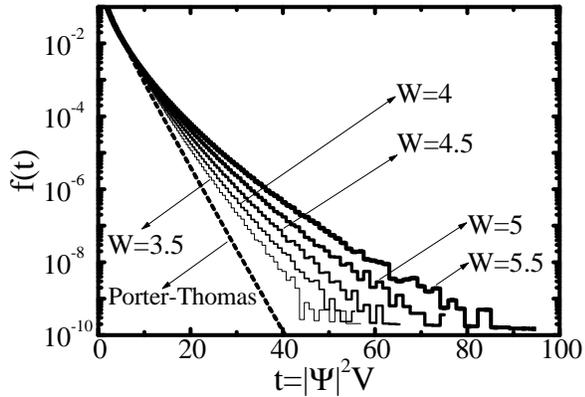,height=3.0in,angle=-90} }
\vspace{0.2in} \caption{Statistics of eigenfunction
``intensities'' $|\Psi_E({\bf m})|^2$ in the band center $E \in
(-0.15,0.15)$ of an Anderson model on a simple cubic lattice
$12^3$. Each curve is obtained by examining about two million
exact eigenstates of the (time-reversal and spin-rotation
invariant) Hamiltonians whose on-site potential random variable
uniformly distributed over the interval $[-W/2,W/2]$. The disorder
strengths $W$ for the five different sets of 30000 Hamiltonians are
chosen to ensure the diffusive ($L \gg \ell$) and semiclassical
($k_F\ell > 1$) transport regime in the conductors they model
(relevant transport quantities are listed in Table~\ref{tab:g}.)
The random matrix theory prediction for the limit $g \rightarrow \infty$
is Porter-Thomas distribution, plotted here as a reference.}
\label{fig:principal}
\end{figure}

\begin{figure}
\centerline{ \psfig{file=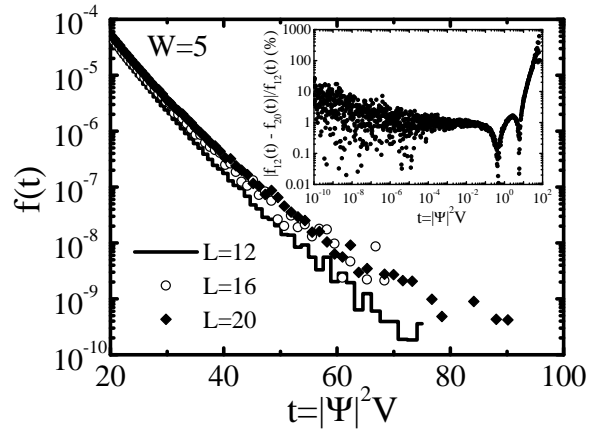,height=3.0in,angle=-90} }
\vspace{0.2in} \caption{Comparison of the $W=5$ eigenstate
statistics $f(t)$ from Fig.~\ref{fig:principal} for a lattice
$12^3$ and the statistics of band center eigenstates generated in
the ensembles of 2000 or 500 conductors modeled on larger
lattices $16^3$ (with $\simeq 150$ eigenfunctions picked in
a small energy interval $\Delta E=0.3$ around  $E=0$ in each
sample) or $20^3$ (with $\simeq 300$ eigenfunctions investigated
in each sample), respectively.} \label{fig:largecube}
\end{figure}

Statistical properties of eigenstates in a closed sample are
described by a disorder-averaged
distribution~\cite{efetov,fmmodes} of eigenfunction
``intensities'' $|\Psi_\alpha({\bf r})|^2$
\begin{equation}\label{eq:ft}
    f(t)=\frac{1}{\rho(E) N} \left \langle \sum_{{\bf r},\alpha}
    \delta(t-|\Psi_\alpha({\bf r})|^2 V) \delta(E-E_\alpha)
  \right \rangle,
\end{equation}
on $N$ discrete points ${\bf r}$ inside a sample of volume $V$.
Here $\rho(E) = \langle \sum_\alpha \delta(E-E_\alpha) \rangle$
is the mean level density at energy $E$, and $\langle \ldots
\rangle$ denotes disorder-averaging. Normalization of eigenstates
gives $\bar{t} = \int d t \, t \, f(t)=1$. We evaluate this
function for eigenstates in the band center $E=0$, which are
obtained by exact numerical diagonalization of the
Hamiltonian~(\ref{eq:tbh}) [note that energy is a parameter in
$f(t)$]. The distribution functions $f(t)$ for the five sets of
conductors modeled on the lattice $12^3$ is shown in
Fig.~\ref{fig:principal}. The explicit dependence of $f(t)$ on
the sample size is demonstrated in Fig.~\ref{fig:largecube} where
$W=5$ case is studied also on the $16^3$ and $20^3$ lattices. By
searching through many configurations of the random potential one
can find the rare ones which are responsible for the appearance
of states with the highest possible amplitude splashes. We smooth
out the data by additional averaging over a small energy interval
(which taken alone, or combined with only a small number of
disorder realizations, is not enough to study the prelocalized
states), without introducing any artifact in the computed
distribution functions.~\cite{bkn} This finally  brings the number
of analyzed eigenstates to about $2 \cdot 10^6$ for each curve
plotted in Fig.~\ref{fig:principal}.

The rest of the paper is organized as follows. In
Sec.~\ref{sec:qchaos} we provide a short survey of the main
analytical results for the eigenstate statistics in 3D, which are
expected to be relevant for our observations. Then in
Sec.~\ref{sec:fit}, a detailed comparison between these
predictions, for both small and large deviations of distribution
functions from RMT, and our results is undertaken using the
calculated transport properties of the samples. Finally, we
conclude in Sec.~\ref{sec:conclusion} by looking beyond
the raw numbers and pointing out at open questions.

\section{Statistical approaches to nonintegrable quantum
systems}\label{sec:qchaos}

Quantum dynamics of a non-interacting particle in random
potential (e.g., generated by quenched impurities) has a long
history of being a standard playground for the development of
ideas of Anderson localization~\cite{anderson} and mesoscopic
physics.~\cite{mqp} The classical counterpart of this problem is
obviously chaotic, but it is only over the past two decades that
its connections~\cite{efetovbook,ghur} to generic clean (i.e.,
without stochastic disorder) examples of quantum
chaos~\cite{chaos} have been deepened. In both quantum chaotic
and quantum disordered systems eigenstates are characterized
solely by their energy, rather than by a set of quantum numbers.
Since eigenstates and eigenvalues cannot be obtained
analytically, it becomes useful to resort to some statistical
treatment where one studies correlators averaged over large
number of eigenstates instead of focusing on the properties of a
single quantum state. Although the methods (and the language) of
quantum chaos and quantum disorder have evolved independently, it
was realized that Wigner-Dyson (WD) level statistics of
RMT,~\cite{ghur} a fingerprint~\cite{bohigas,andreev} of quantum
chaotic systems, is also applicable to disordered
systems.~\cite{gorkov,efetovbook} However, the lack of any
transport-related energy or time scale in RMT description of the
delocalized phase signals that relevant time to traverse the
sample diffusively, $t_{\mathcal D} = L^2/{\mathcal D}$
(${\mathcal D}=v_F\ell/3$ being the bare classical diffusion
constant in 3D), is set to zero in this framework. Therefore, it
became clear that WD statistics can be applicable to disordered
system spectra only for the energy separation scale much smaller
than the Thouless energy~\cite{shklovskii} $E_{\rm
Th}=\hbar/t_{\mathcal D}$.

The physical origin of deviations from RMT in disordered systems 
with $g < \infty$ is the finite time $t_{\mathcal
D}$ required for the particle to spread ergodically all over the
sample, i.e., for the classical motion to explore the whole phase
space. The statistical approaches to quantum disorder problems,
like supersymmetric nonlinear $\sigma$-model
(NLSM)~\cite{efetovbook} which maps stochastic problem to a field
theory without randomness, provide the justification of RMT in
the ergodic ($t \gg t_{\mathcal D}$) regime of diffusive
dynamics. Furthermore, these techniques make it possible to study
also the deviations from RMT for non-ergodic times or energy
scales in weakly disordered ($k_F\ell \gg 1$) conductors---both
perturbative and nonperturbative corrections are governed, within
this framework, by the diffusion operator describing the dynamics
of a corresponding classical system. The well-understood quantum
chaotic properties of disordered system have made these systems a
standard laboratory to test different approaches to generic
quantum chaos~\cite{andreev,smith} (where well-defined averaging
procedure over an ensemble is lacking~\cite{prange}). Thus,
limitations of RMT encountered in disordered electron systems
have put the study of deviations from the universality regime
into the focus of both mesoscopic and quantum chaos
communities,~\cite{mirlin,mirlinbal} where ``lessons from
disordered metals''~\cite{simons} have greatly influenced the
development of formalism for more arduous examples of quantum
chaos.

Recently, the equivalent program has been pursued for the
eigenfunction statistics.~\cite{mirlin} The earliest prediction
for the (universal) distribution of eigenfunction intensities, in
Gaussian orthogonal ensemble (GOE) of random matrices (which are
time-reversal and spin-rotation invariant),
\begin{equation}\label{eq:porter}
  f_{\text{PT}}(t)=\frac{1}{\sqrt{2 \pi t}} \exp(-t/2).
\end{equation}
is known as the Porther-Thomas~\cite{porter} (PT) distribution.
Assuming only that eigenfunctions are normalized but otherwise
arbitrary, $f_{\text{PT}}(t)$ can be derived~\cite{haake} from
the probability that, e.g., component $\Psi_1$ [which corresponds
to $\Psi({\bf m}_0)$ at some point ${\bf m}_0$ inside the sample]
of an eigenstate of a $N \times N$ random matrix is equal to some
value $t/N$
\begin{eqnarray}\label{eq:ptsmoly}
\lefteqn{f_{\rm PT}(t)  =  \lim\limits_{N \rightarrow \infty}
\frac{1}{N\sqrt{\pi}} \frac{\Gamma(N/2)}{\Gamma[(N-1)/2]}}
\nonumber \\
&& \mbox{} \times \int\limits_{-\infty}^{\infty} \,
[\prod_{i=1}^N d\Psi_i] \,  \delta (t-N|\Psi_1|^2)  \, \delta
(1-\sum_{j=1}^N|\Psi_j|^2)  \nonumber \\
&& \mbox{} = \lim\limits_{N \rightarrow \infty}
 \frac{1}{\sqrt{N \pi t}} \frac{\Gamma(N/2)}{\Gamma[(N-1)/2]}
 \left(1-\frac{t}{N} \right )^{\frac{N-3}{2}},
\end{eqnarray}
in the limit $N \rightarrow \infty$ (which corresponds to an 
infinitely large lattice in our problem). This is essentially an
example of the central-limit theorem, and it should describe
completely the eigenstate statistics in the universality limit
that is insensitive to any physical details of the system. Such 
limit in disordered electron systems requires infinite 
conductance $g = E_{\rm Th}/\Delta \rightarrow \infty$ since 
level spacing $\Delta$ (thermodynamic scale) sets the smallest 
energy scale and $t_{\mathcal D} \rightarrow 0 \Leftrightarrow 
E_{\rm Th} \rightarrow \infty$ (in real systems
$E_{\rm Th}$ is large only in small enough samples, such as
quantum dots). The universality stems from the basis invariance of
RMT, i.e., the fact that eigenfunctions in RMT are structureless
with $\Psi_\alpha({\bf r})$ and $\Psi_\alpha({\bf r}^\prime)$
being uncorrelated for $|{\bf r}-{\bf r}^\prime| \gtrsim \ell$,
and fluctuating just as Gaussian random variables. However, random
Hamiltonians of real disordered solids are tied to a real-space
representation, where matrix elements are spatially dependent and
TBH~(\ref{eq:tbh}) is a band diagonal matrix. Therefore, they do
not satisfy  statistical assumptions of the standard RMT
ensembles since all elements of such random matrices are non-zero
and spatially independent. Nevertheless, a rigorous connection to
the RMT eigenstate statistics [here just heuristically 
established through the interpretation of Eq.~(\ref{eq:ptsmoly})] is
provided by Efetov's supersymmetric
approach~\cite{efetov,efetovbook} (i.e., zero-dimensional limit of 
the NLSM). While WD statistics works well
for the part of spectrum contained within the interval
$|E-E^\prime| \ll E_{\rm Th}$, the distribution function $f(t)$
in finite $g$ systems (like the ones in
Fig.~\ref{fig:principal}) do not overlap with PT distribution in
any interval of eigenfunction amplitudes. The redistribution of
amplitude statistics, caused in part by the appearance of highly
unlikely according to RMT prelocalized states, leads to three
different regions of intensities $t$. The deviations are the
strongest in the large-$t$ limit where $f(t)$ can be orders of
magnitude greater than PT distribution. This occurs also in
quantum chaos~\cite{kaplan} where localization due to scars is
generally less pronounced than inhomogeneities of the
prelocalized states in strong enough disorder.~\cite{sridhar}

Small deviations of $f(t)$ from the PT distribution are accounted
by a WL-type correction~\cite{fyodorov} (i.e., a quantity
sensitive to the breaking of time-reversal symmetry; here we use
the expression for GOE)
\begin{equation}\label{eq:fm}
  f_{\rm FM}(t)=f_{\rm PT}(t)\left[1+\frac{\kappa}{2}
  \left(\frac{3}{2}-3t+\frac{t^2}{2}\right) + {\mathcal O}(g^{-2})
  \right].
\end{equation}
which is a regular function in the small parameter $1/g$ and can
be derived by a perturbative treatment of the nonzero spatial
modes of the NLSM. This result was obtained by Fyodorov and
Mirlin~\cite{fyodorov} (FM) for $t \ll \kappa^{-1/2}$ (where
correction is smaller than the RMT term). As in the case of
corrections to WD statistics, the deviations are parameterized by
the properties of classical diffusive dynamics. Namely, $\kappa$
is defined in terms of the diffusion propagator (i.e.,
one-diffuson loop)~\cite{mirlin}
\begin{equation}\label{eq:kappa}
  \kappa_{\rm diff} \equiv \Pi({\bf r},{\bf r}) = \frac{2}{g \pi^2}
  \, \sum\limits_{\bf q} \frac{\exp(-{\mathcal D}{\bf q}^2\tau)}{{\bf
  q}^2L^2},
\end{equation}
which is the sum over the diffusion modes (diffusion propagator
$\Pi({\bf r},{\bf r}^\prime)$ is the Green function of the
diffusion equation and can be expressed in terms of eigenvectors
and eigenvalues of the diffusion operator $-{\mathcal D}{\bf
q}^2$, in a rectangular geometry studied here). The sum in
Eq.~(\ref{eq:kappa}) diverges linearly in 3D at large momenta,
thus requiring a cutoff at $|{\bf q}| \sim 1/\ell$ to retain the
validity of the diffusive approximation. We provide the
ultraviolet regularization using exponential damping
factor~\cite{montsum} which limits the sum to the diffusive regime
${\mathcal D} {\bf q}^2 \ll 1/\tau$ ($\tau=\ell/v_F$ is the
elastic mean free time)
\begin{equation}\label{eq:kappasum}
 S(y)= \frac{1}{4}\sum_{n_x,n_y,n_z \neq 0}\,\frac{\exp[-4\pi^2(n_x^2+n_y^2+n_z^2)\,y]}
  {n_x^2+n_y^2+n_z^2}.
\end{equation}
Here the wave vectors ${\bf q}$ are quantized by the periodic
boundary conditions used in all directions, $q_x=2\pi n_x/L$,
$n_x=\pm 1, \pm 2, \ldots$, and correspondingly for $q_y$ and
$q_z$. The argument $y$ is expressed in terms of the semiclassical
transport quantities
\begin{equation}\label{eq:y}
  y=\frac{\mathcal D \tau}{L^2}=\frac{1}{3} \left (\frac{\ell}{L}
  \right)^2.
\end{equation}
The sum~(\ref{eq:kappasum}) can be evaluated exactly by a simple
numerical computation. On the other hand, its analytical dependence
on $\ell/L$ is usually obtained after approximating the discrete
summation by an integral (a standard result in the literature is
therefore quoted~\cite{fyodorov} as $S(y)\sim L/\ell$). To avoid
loosing the terms of the original sum, which  depend on specific boundary
conditions,~\cite{montsum} it is necessary to evaluate the
original discrete form. Following Ref.~\onlinecite{montsum} this
can be done by using the function $F(y)=\sum_{n=1}^{\infty} \,
\exp(-\pi^2n^2y)$,
\begin{equation}\label{eq:sdiff}
  \frac{\partial S(y)}{\partial y} = -\pi^2 [2F(4y)]^3.
\end{equation}
Since $F(y)$ is related to the complete elliptic integrals, for
small values of argument $y \ll 1$ they can be approximated by the
leading order to give
\begin{equation}\label{eq:felliptic}
    F(y) \simeq \frac{1}{2}\left[(\pi y)^{-\frac{1}{2}}-1 \right],
\end{equation}
which, upon integration in~(\ref{eq:sdiff}), leads to
\begin{equation}\label{eq:sumabc}
  S(y)=\frac {\sqrt{\pi}}{4\sqrt {y}} +\frac{3}{4}\,\pi \,\ln (y)-3\,\pi^{3/2} \,
\sqrt {y}+{\pi }^{2}y + \alpha_P.
\end{equation}
The integration constant $\alpha_P$ can be fixed numerically by
finding the limit, as $y \rightarrow 0$, of the difference between
exactly calculated discrete sum $S(y)$ and its analytical
approximation~(\ref{eq:sumabc}) with removed constant $\alpha_P$. We find that
this difference converges at small $y<10^{-4}$ to $\alpha_P
\approx 8.32$. For the largest $\ell$ in our study, the function
(\ref{eq:sumabc}) with this $\alpha_P$ reproduces the numerically
calculated sum over the diffusion modes to within $2\%$.
Evaluation of $S(y)$ in the standard way, by approximating it
with an integral, gives only the leading term (which is the
same for all boundary conditions)
\begin{equation}\label{eq:sumaint}
  S_I(y)=\frac {\sqrt{\pi}}{4\sqrt {y}} - \alpha_I,
\end{equation}
where $\alpha_I$ is an integration constant defined by the
infrared cutoff at small wave vectors $|{\bf q}|\sim L^{-1}$.

The divergence of the sum over diffusion modes~(\ref{eq:kappasum})
in 3D points out that more careful treatment is needed of the
short-scale physics.  Since $\kappa$ has the meaning of a
time-integrated return probability for a diffusive
particle,~\cite{mirlin} it can be generalized to the ballistic
case (various transport and thermodynamic phenomena encountered
in disordered conductors are related to the classical return
probabilities for a diffusive particle, see
Ref.~\onlinecite{monty}). Using ballistic
generalization~\cite{balsigma} of the NLSM to go beyond the
diffusive approximation, it was shown~\cite{mirlin,mirlinbal} that
corresponding ballistic contribution to this return probability,
i.e., the probability for a particle to be scattered only once
from an impurity and return back after a time $t \ll \tau$, has
to be added to $\kappa_{\rm diff}$. Therefore, the total return
probability is expected to be given by $\kappa=\kappa_{\rm
diff}+\kappa_{\rm bal}$. Such (semiclassical) ballistic effects
are non-universal, i.e., they can depend strongly on the microscopic
details of disorder (they are negligible in the case of smooth
potential having a correlation length much bigger than~\cite{mirlin,uski2001}
$1/k_F$). In fact, it was shown recently~\cite{bkn}  that
$\kappa_{\rm bal}$ determining perturbative corrections in the
Anderson model can be much greater than $\kappa_{\rm diff}$. In
Sec.~\ref{sec:fit} we show explicitly that both contributions
are needed to describe $\kappa$ as a function of disorder strength,
contrary to some previously reported results~\cite{uski} for the
3D Anderson model where fitted $\kappa$ was related to ballistic
effects only.

The second point of intersection of computed distribution function
with PT distribution can be considered as the point at which a
tail of $f(t)$ starts to form. While $f_{\rm FM}(t)$ still
approximately describes the ``near tail'' around these values of
$t$, the slow decay of $f(t)$ eventually requires a
nonperturbative formula in $1/g$. Such is provided by the 3D case
calculations~\cite{mirlinprb} in the framework of standard
diffusive NLSM, which give with an exponential accuracy
\begin{equation}\label{eq:nlsm}
f_{\rm NLSM}(t) \sim \exp \left [- \frac{1}{4 \kappa} \, \ln^3
(\kappa t) \right ],\ t \gtrsim \kappa^{-1},
\end{equation}
and stems from the appearance of the prelocalized states. Since $\kappa \sim
(k_F\ell)^{-2}$ (only in the leading order---the discrete sum in
Eq.~(\ref{eq:kappasum}) also contains other terms), this result
written down to the leading order term of the cubic polynomial in
$\ln[t/(k_F\ell)^2]$,
\begin{equation}\label{eq:nlsm1}
-\ln f_{\rm NLSM}(t) \sim (k_F\ell)^2 \ln^3 \left
[\frac{t}{(k_F\ell)^2} \right ],
\end{equation}
emphasizes that prefactor contains $(k_F\ell)^2$ dependence on
the disorder strength. However, the numerical constant is
uncertain~\cite{mirlinprb} because of being determined by the
ultraviolet ${\bf q}$ (in the field theoretical language), i.e.,
length scales $\lesssim \ell$ which are outside of the diffusive
NLSM framework. Namely, the standard nonlinear
$\sigma$-models~\cite{efetovbook,lerner} are long-wavelength
effective field theories for the diffusive modes whose saddle
point is analyzed to get the eigenstate
statics.~\cite{muz,efetov,mirlinprb} The role of energy is played
by the diffusion operator $-{\mathcal D}\nabla^2$, so that
fundamental variables of the theory, $Q({\bf r})$ matrix fields,
should vary spatially on the scale much larger than $\ell$.
However, it was found~\cite{efetov,mirlinprb} that in 3D systems
$Q({\bf r})$ vary rapidly on the length scale $\sim \ell$, which
then impedes the prospect of getting rigorous results and points
out to a different physics determining the structure of
eigenfunctions in 3D, than is the case if low-dimensional
systems.~\cite{mirlin,uskiprb} An attempt to overcome this
limitations, by using ballistic NLSM which extends the semiclassical
description to all momenta $|{\bf q}| \lesssim k_F$, leads to the same
$(k_F\ell)^2$ prefactor dependence but offers $\pi/9\sqrt{3}$ as
the precise value of its constant piece.~\cite{dek}

In between the perturbative ($t \ll \kappa^{-1/2}$) and the
far-tail region ($t > \kappa^{-1}$) of the wave function
amplitudes, NLSM analysis of $f(t)$ predicts an intermediate
range of amplitudes, described by~\cite{efetov}
\begin{eqnarray}\label{eq:interm}
f_{\rm IM}(t) & \simeq & \frac{1}{\sqrt{2 \pi t}}\exp \left[
\frac{1}{2} \left(-t+\frac{\kappa t^2}{2} + \cdots \right) \right
], \nonumber \\
&& \kappa^{-1/2} \lesssim t \lesssim \kappa^{-1}.
\end{eqnarray}
This has also has the form of the corrected PT distribution, as is
the case of FM distribution. However, the correction term here is
in the exponent, and therefore of a different type than the one
in $f_{\rm FM}(t)$. It should be large compared to unity, but
small compared to the leading RMT term.~\cite{mirlin}
\begin{table}
\begin{center}
\begin{tabular}{|c|c|c|c|}
      &  $g$ & $\ell \ (a) $ & $\kappa_{\rm diff}$ \\ \tableline
  $W=3.5$ & $21.5$ & $2.89$  &   0.00074        \\
  $W=4.0$ & $17.4$ & $2.21$  &   0.0031        \\
  $W=4.5$ & $14.3$ & $1.75$  &   0.0084          \\
  $W=5.0$ & $11.9$ & $1.42$  &   0.018        \\
  $W=5.5$ & $10.0$ & $1.17$  &   0.035      \\
\end{tabular}
\end{center}
\vspace*{0.2cm} \caption{Transport properties computed for our
five ensembles of impurity configurations characterized by
different diagonal disorder strength $W$ in the Anderson model on
a simple cubic lattice $12^3$.} \label{tab:g}
\end{table}

An alternate route to account for the slow decay of $f(t)$ at
large amplitudes has been undertaken through a ``direct optimal
fluctuation method'' (DOF) of Ref.~\onlinecite{smolyarenko}. This
approach suggests possible importance of ballistic
non-semiclassical effects, which are missed in both diffusive and
ballistic NLSM formulation, by analyzing the short-scale
structure of solutions of the Schr\" odinger equation (i.e., by
analyzing the saddle point of the original problem of quantum
particle in a random potential, rather than the saddle point of
its effective field theory obtained by integrating out the
disorder degrees of freedom). It gives the following result
for the far tail asymptotics (where only the leading log-cube
term of the full cubic polynomial of $\ln t$ is computed
explicitly)
\begin{equation}\label{eq:smoly}
f_{\rm DOF}(t) \sim \exp \left [- C_{\rm DOF} \, k_F\ell \,
\ln^3 t \right ],
\end{equation}
assuming Gaussian white-noise random potential, and estimating
$C_{\rm DOF} \simeq 3 \cdot 10^{-3}$ for that type of disorder.
The dependence on the prefactor on $k_F\ell$ here is linear, which
can substantially increase the probability to observe a rare
event as compared to NLSM prediction. Nonetheless, this result
has also been interpreted heuristically within the NLSM
picture,~\cite{mirlin} assuming that  $\kappa$ might depend solely
on the non-universal (semiclassical) ballistic $\sim 1/k_F\ell$
contributions of the same kind as those encountered in the region
of small eigenfunction amplitudes.

\section{Fitting the eigenstate statistics: physics behind deviations
from random matrix theory} \label{sec:fit}

In this section, we first describe the details of the computation
of $f(t)$, and also give elementary account of the quantum
transport properties of conductors described by the Anderson
model. This should serve as an overture for the subsequent
detailed examination of the eigenstate statistics in the
perturbative region (i.e., the main body of the distribution
function of intensities), and nonperturbative region (where
amplitude splashes generate large deviation from PT
distribution), by attempting to fit the analytical forms introduced in
Sec.~\ref{sec:qchaos}. We then analyze the confidence in such
descriptions through relative error of the corresponding fits.
The extracted fitting parameters are interpreted by comparing
them to the expected ones in the analytical predictions, with the
help of transport properties of our finite-size samples.

We solve exactly the eigenproblem of the TBH~(\ref{eq:tbh})
of a finite-size system inside  a small energy window $\Delta
E=0.3$ positioned around $E=0$. This interval picks 60--70 states
in each of the 30000 conductors (modeled on the lattice $12^3$) with different
impurity configurations. The ensemble of disordered samples is
characterized by the disorder-averaged conductance $g$. The
overall number of collected states depends on disorder---as the
disorder is increased the energy band broadens, meaning that some
states start to appear with energy eigenvalues beyond the band
edge $E_b=6t$ of the clean TBH while the average number of states in the band
center decreases. These states are used to evaluate $f(t)$ as a
histogram of intensities at all points inside the sample
($N=12^3$). The two delta functions in Eq.~(\ref{eq:ft}) are
approximated by box functions $\bar{\delta}(x)$. The width
$\Delta E=0.3$ of $\bar{\delta}(E)$ is such that $\rho(E)$ is
constant inside $\Delta E$. The amplitudes of wave functions are
sorted in the bins defined by $\bar{\delta}(t-|\Psi_\alpha({\bf
r})|^2 V)$, where their width is constant on a logarithmic scale.

As emphasized in Sec.~\ref{sec:intro}, before embarking on the
search for prelocalized states, we first compute the transport
properties of our samples. The exact zero-temperature conductance
of the nanoscale finite-size sample is obtained from the
Landauer-type formula~\cite{lb}
\begin{equation}\label{eq:landauer}
  g = \text{Tr} \, \left [ {\bf t}(E_F) {\bf t}^{\dag}(E_F) \right ],
\end{equation}
where transmission matrix ${\bf t}(E_F)$ is expressed in terms of
the real-space Green functions~\cite{datta} for the sample
attached to two disorder-free semi-infinite leads. The details of
such calculations for the lattice model studied here are given
elsewhere.~\cite{allen} Here we just clarify the relationship of
this conductance to the Thouless conductance $g_{\rm Th}=2\pi
E_{\rm Th}/\Delta$ (expressed in terms of the spectral properties
of a closed sample), which appears in standard analytical
treatments of various disordered problems, including the
eigenstate statistics.~\cite{mirlin} This mesoscopic
computational technique opens the sample to the surrounding
ideally conducting medium, so that particles can leave and enter
through the lead-sample interface. Thus, the discrete levels of
an initially isolated sample are smeared, and spectrum of {\it
sample+leads=infinite system} becomes continuous (which allows us
to find the conductance at any $E_F$ inside the band). However,
the computed conductance, for not too small
disorder~\cite{mackinnon,nikolic_qpc} or coupling to the leads of
the same transverse width as the sample~\cite{mackinnon}, is
practically equal to the ``intrinsic'' conductance $g_{\rm Th}$.
In practice this means that we attach sample to the leads of the
same cross section and use the same hopping matrix element
$t_{{\bf m} {\bf n}}$ throughout the system (i.e., in the leads,
lead-sample coupling and in the sample), in order to minimize any
influence which leads can have on the
conductance.~\cite{nikolic_qpc}

The disorder-averaged transport quantities, characterizing the
five ensembles of conductors studied here, are listed in
Table~\ref{tab:g}. For weak disorder $W \lesssim 6$ conductance
$g$ is dominated by the semiclassical effects.~\cite{allen} Thus,
we use the Bloch-Boltzmann formalism (applicable when~\cite{allen}
$\ell \gg a$), in Born approximation for the scattering on a single
impurity, to get the elastic mean free path $\ell(E_F=0) \simeq 35.4/W^2$
shown in Table~\ref{tab:g}. Analytical treatments usually assume a
simple spherical Fermi surface for which  $k_F$ is just the
radius of the sphere. Such quantity is not well-defined for a
lattice system with non-spherical Fermi surface where $k_F$ is
direction dependent (i.e., different average values can be
obtained depending on whether one averages the absolute value of
$k_F$ or the root-mean-square of $k_F$ over the Fermi surface ).
Nevertheless, all different averaging procedures give similar
values, and we use conventionally the one which would reproduce
some transport formula, like Sharvin~\cite{sharvin} classical
point contact conductance $G=G_Q\, k_F^2 L^2/4\pi$, where such
average values (here over the Fermi surface $E_F=0$ of a simple
cubic lattice) naturally appear. This convention gives $k_F
\approx 2.8/a$, which should serve as a counterpart of $k_F$
appearing in theoretical simplifications assuming Fermi sphere.
It is easy to check that these values of parameters, plugged into
the Drude-Boltzmann formula $g \simeq (k_F^2 L^2/4 \pi) \,
(\ell/L)$, approximately reproduce the disorder-averaged Landauer
two-probe conductances from Table~\ref{tab:g}.

\subsection{Perturbative deviations from the RMT limit}

\begin{figure}
\centerline{ \psfig{file=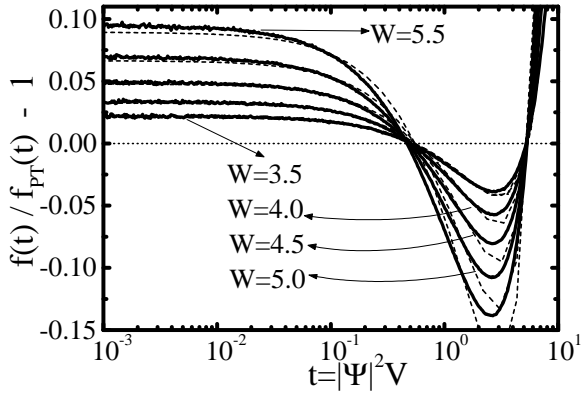,height=3.0in,angle=-90} }
\vspace{0.2in} \caption{Weak localization correction $\kappa/2\,
[3/2-3t+t^2/2]$, through which $f_{\rm FM}(t)$ accounts for the
small deviations from the Porter-Thomas distribution $f_{\rm
PT}(t)$, fitted (dashed lines) to $f(t)/f_{\rm PT}-1$ in the
perturbative region of amplitudes for statistics from
Fig.~\ref{fig:principal}. The first zero of all curves is $W$
dependent and falls in the interval (0.46,0.51); the second one
is at $\approx 0.53$. This should be compared to the two zeros
$t_0^\prime=0.55$ and $t_0^{\prime\prime}=5.45$ of
$3/2-3t+t^2/2$. This fit is expected to be valid from
$t<1/\sqrt{\kappa}$ to $t=0$---we find the relative error $[f(t)-
f_{\rm FM}(t)]/f(t)$ to be less than 1\% all the way to $t \sim
10^{-11}$ (up to some noise in the data at the smallest
investigated $t$).} \label{fig:wl}
\end{figure}

We commence the comparison between our results in
Fig.~\ref{fig:principal} and functional forms from
Sec.~\ref{sec:qchaos} by fitting FM distribution ~(\ref{eq:fm})
to the data in the region of amplitudes where $f(t)$ deviates
only slightly from PT distribution. The plot of $f(t)/f_{\rm
PT}(t)-1$ (Fig.~\ref{fig:wl}) offers a straightforward way to
check for the non-trivial feature of the correction term, such as
whether the zeros $t_0^\prime=0.55$ and $t_0^{\prime\prime}=5.45$
of $3/2-3t+t^2/2$ are exhibited by our data. We find that zeroes
of the curves from our numerical simulation are slightly smaller
(Fig.~\ref{fig:wl}), the first one being even disorder strength
dependent. Since naive (visual) inspection of fits, especially on
a logarithmic scale might lead to erroneous conclusions (like the
range of amplitudes where some function fits the data), we
establish a quantitative criterion of the quality of a fit with
some formula $f_{\rm fit}(t)$ by looking at the relative error
$[f(t)-f_{\rm fit}(t)]/f(t)$. This becomes especially important in
assessing the fit of the far tail on the semilogarithmic scale (an
example of such assessment is shown in Fig.
\ref{fig:procent_elc}). Moreover, this type of plot directly
highlights intervals where some analytical formula can be
considered to describe the data. For example, this gives a
quantitative insight into the boundaries of perturbative,
intermediate and nonperturbative regions discussed in
Sec.~\ref{sec:qchaos}. The FM distribution describes $f(t)$ for
small eigenfunction amplitudes completely in the weakly
disordered samples, where $[f(t)-f_{\rm FM}(t)]/f(t)$ is less
than 1\% for the data and the fit in Fig.~\ref{fig:wl}. It also
fits the statistics in the conductors where dirty metal regime of
transport is reached in the Anderson model (e.g., the resistivity
of $W=5.5$ sample would be $\simeq 500\, \mu \Omega$cm for
lattice spacing $a=3$ \AA). However, it is demonstrated by
Figs.~\ref{fig:wl} that increasing $W$ toward the boundary $W
\simeq 6$ leads to larger relative difference between the fitted
FM distribution function and the numerical data, while also
shrinking the interval of $t$ where such comparison is still
reasonable in the first place. For $W \gtrsim 6$ unwarranted
application of the Drude-Boltzmann formula would give $\ell < a$
(i.e., the transport in this regime is dominated by
non-semiclassical and nonperturbative effects~\cite{allen}).
Therefore, even though it might be possible to claim that FM form
accounts for some portion of $f(t)$ at stronger
disorder,~\cite{uski} the direct comparison of extracted
parameter $\kappa$ to the calculated one becomes nonsensical. For
example, $W$ cannot be interpreted as $\sim \ell^{-1/2}$ in this
regime. In fact, the departure of FM correction from our $f(t)$
starts before the upper limit of disorder, determining the
breakdown of Bloch-Boltzmann theory of semiclassical transport, is
reached. Similar distinction between the disorder limitation for
the diagrammatic results and somewhat higher limit for the
validity of Boltzmann equation is seen in some other
cases~\cite{kubo} (despite the fact that Boltzmann theory is
rigorously justified as a lowest order result of a perturbative
expansion of e.g., the Kubo formula).

\begin{figure}
\centerline{ \psfig{file=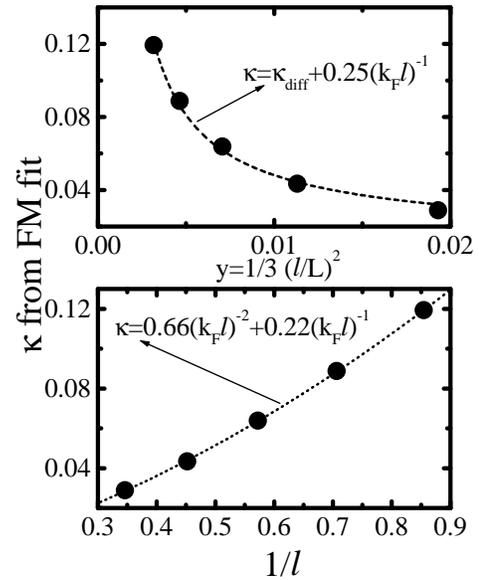,height=3.0in,angle=0} }
\vspace{0.2in} \caption{The extracted $\kappa$ from the fits of
FM distribution (Fig.~\ref{fig:wl}) to the portion of our
numerical distribution $f(t)$ contained within the interval of
$t$ where deviations from the Porter-Thomas distribution of RMT
are small. The dependence of these values on the disorder strength
(i.e., mean free path $\ell$) is explained as a sum of the
diffusive contribution $\sim 1/(k_F\ell)^2$ (or, more precisely,
$k_{\rm diff}$ expressed through the full sum~(\ref{eq:sumabc}),
as shown in the upper panel) and the ballistic contribution $\sim
1/k_F\ell$.} \label{fig:kappa}
\end{figure}

The one-parameter fit of the FM result Eq.~\ref{eq:fm} allows us
to extract the first relevant physical parameter from our data, $\kappa$.
The extracted values are presented in Fig.~\ref{fig:kappa}. These values 
of $\kappa$ do not match to the calculated $\kappa_{\rm diff}$  (Table~\ref{tab:g}) determined by
the universal properties of the diffusive dynamics. This is not surprising in the light of the
discussion in Sec.~\ref{sec:qchaos}. What might be surprising is that
disorder-specific $\kappa_{\rm bal}=\kappa-\kappa_{\rm diff}$ can
be few times greater than $\kappa_{\rm diff}$. This points out to
the importance of the short-scale effects even in the
perturbative region. Nevertheless, it appears that non-Gaussian
features (non-zero higher order cumulants) of the uniformly 
distributed random potential of the Anderson model do not 
generate interesting terms beyond the second-order ones 
computed explicitly in the FM correction.

For a system of fixed size, Figure~\ref{fig:kappa} shows that
$\kappa$ changes with the strength of disorder in a way expected
for the change of a sum of the two contributions. The first term
in the phenomenological formula for $\kappa =0.66(k_F\ell)^{-2} +
0.22(k_F\ell)^{-1}$ is different from the analytically computed
$\kappa_{\rm diff}$. However, the function fitting the data  on
the lower panel of Fig.~\ref{fig:kappa} is only the leading order
behavior (usually given when boundary conditions quantizing the
diffusion modes are neglected~\cite{montsum}) of the full
$\kappa_{\rm diff}$ computed in Sec.~\ref{sec:qchaos}. Therefore,
we proceed to fit formula~(\ref{eq:kappa}) in the following form
\begin{equation}\label{eq:fullkappa}
  \kappa=\frac{K_1}{\sqrt{y}}
  \frac{2S(y)}{\pi^2}+\frac{K_2}{\sqrt{y}}.
\end{equation}
That such fit of $\kappa$ vs $y=1/3 (\ell/L)^2$ is successful is
shown on the upper panel of Fig.~\ref{fig:kappa}, and also by
comparing $K_1/\sqrt{y} =0.012L/\ell$ factor in the first term to
the expected $1/g=(4\pi/k_F^2L^2)(L/\ell)=0.011$. The remarkably
close numerical values confirm suggest that $\kappa_{\rm diff}+0.23(k_F\ell)^{-1}$
should include the lower order boundary condition dependent terms~\cite{montsum}
in $\kappa_{\rm diff}$. While it would be hard to investigate this extra
terms from the scaling of numerically computed disorder-averaged conductance
(because of ``pollution'' by the conductance fluctuations), here they
parameterize the features of mesoscopic fluctuations themselves.

\begin{figure}
\centerline{ \psfig{file=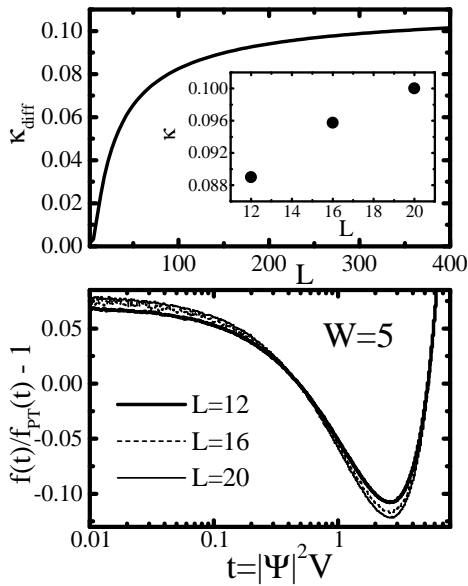,height=3.0in,angle=0} }
\vspace{0.2in} \caption{The region of small deviations of $f(t)$
from the RMT statistics $f_{\rm PT}(t)$ in the three sets of
$W=5$ disordered conductors (lower panel), which differ by the
lattice size: $12^3$, $16^3$ and $20^3$
(Fig.~\ref{fig:largecube}). Fitting WL correction $\kappa/2\,
[3/2-3t+t^2/2]$ to these curves gives $\kappa$ as a function of
$L$, as shown in the inset of the upper panel. This should be
compared to size-dependence of its diffusive contribution
$\kappa_{\rm diff}$ in Eq.~(\ref{eq:kappa}), since ballistic
contributions are size independent.} \label{fig:largecube_wl}
\end{figure}

The other important question concerns with the dependence of the
two contributions on the sample size. Standard display of
$\kappa_{\rm diff}$ and $\kappa_{\rm bal}$  in the literature
shows only the leading order terms of such parameters, like
Eq.~(\ref{eq:nlsm1}) or Eq.~(\ref{eq:smoly}), which are
$L$-independent.~\cite{mirlin,smolyarenko} The short-scale
contributions to $\kappa$ are not sensitive to the sample
boundaries, so that change of the fitted $\kappa$ in $f_{\rm
FM}(t)$ with $L$ should be generated solely by a change of
$\kappa_{\rm diff}(L)$. Our  evaluation of $\kappa_{\rm diff}$
introduces several size dependent terms, cumulative effect of
which is shown in the upper panel of Fig.~\ref{fig:largecube_wl}.
In the limit of large system size, this converges asymptotically
to $\lim_{L \rightarrow \infty} \kappa_{\rm diff}(L)=0.11$. This
offers a simple test of the accuracy of the regularization scheme
employed to obtain $k_{\rm diff}(L)$. In order to investigate
this issue, we generate a set of conductors modeled on the
lattices $16^3$ and $20^3$ with $W=5$ disorder strength.
The comparison between the corresponding distributions $f(t)$ is shown
in Fig.~\ref{fig:largecube}; using the same energy window $\Delta
E=0.3$, this distributions are generated from about $\simeq 150$
or $\simeq 300$ states, for each realization of disorder, which
are picked in the band center of $16^3$ or $20^3$ lattice,
respectively. Even without too large statistics, a palpable
deviation between the two distributions is observed in the far
tail, and also in the region of small $t$. In quantitative terms,
analytical expression~(\ref{eq:kappa}) gives $\kappa_{\rm
diff}(L=12)=0.019$, $\kappa_{\rm diff}(L=16)=0.028$, and
$\kappa_{\rm diff}(L=20)=0.036$, while the fitted ones are
$\kappa(L=12)=0.089$, $\kappa(L=16)=0.096$ and
$\kappa(L=20)=0.100$. At lower disorder $W=3.5$ we get somewhat
better agreement between fitted and calculated change in $\kappa$
with the system size: $\kappa(L=20)-\kappa(L=16)=0.0036$ versus
$\kappa_{\rm diff}(L=20)-\kappa_{\rm diff} (L=16)=0.0027$, in
accord with the discussion above on the disorder strength
boundaries for the validity of perturbation theory in $1/k_F\ell$.

\subsection{Far tail}

\begin{figure}
\centerline{ \psfig{file=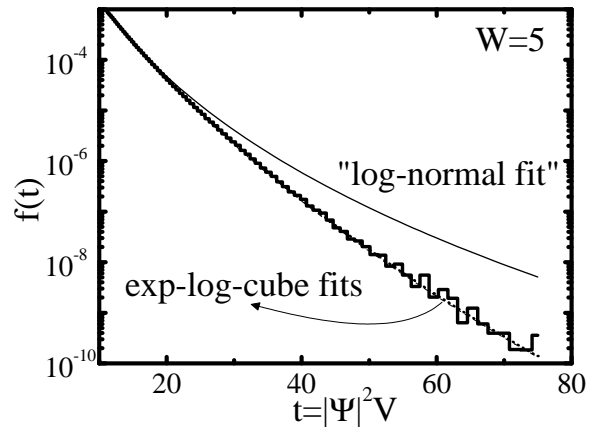,height=3.0in,angle=-90} }
\vspace{0.2in} \caption{Exp-log-cube fits, Eq.~(\ref{eq:elcfit})
and ~(\ref{eq:nlsmfit}), versus attempted log-normal fit to the
far tail of $f(t)$ for $W=5$ ensemble of disordered conductors
from Fig.~\ref{fig:principal}. The numerical data for our
three-dimensional systems clearly favor exp-log-cube asymptotics.
Relative error of the fits is plotted in
Fig.~\ref{fig:procent_elc}.} \label{fig:quest}
\end{figure}

\begin{figure}
\centerline{ \psfig{file=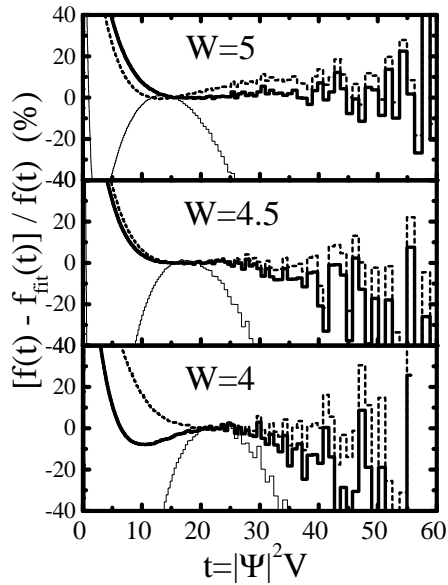,height=3.0in,angle=0} }
\vspace{0.2in} \caption{Relative error of the two different ways of
fitting the exp-log-cube formula (as well as the illustration of
inadequacy of the ``log-normal fit'' from Fig.~\ref{fig:quest},
thin solid line) to the exact $f(t)$ from
Fig.~\ref{fig:principal}: thick solid line is for $f_{\rm fit}(t)
\equiv C_p \exp(-C_3 \ln^3 t)$
 and dashed line is for $f_{\rm fit}^\prime(t) \equiv C_p^{\prime} \exp[-1/4\kappa_{\rm ELC}
\, \ln^3 (\kappa_{\rm ELC} t)]$.} \label{fig:procent_elc}
\end{figure}

A common feature  of (almost) all analytical results for the
eigenstate statistics in 3D metallic samples is that some
exp-log-cube formula is predicted to describe the
large-$t$ asymptotic behavior of $f(t)$. Thus, the most
general functional form along these lines would be an exponent
of a cubic polynomial of $\ln t$. Since lower order terms in the
polynomial are rarely calculated, and fitting of any formula
is more reliable when the number of free parameters is small, we
choose to fit two different simple (employing only two parameters)
exp-log-cube expressions (postponing the physical interpretation
of their parameters for a moment):
\begin{equation}\label{eq:elcfit}
 f_{\rm fit}(t)=C_p \exp(-C_3\ln^3 t),
\end{equation}
which is always the leading order term [and sometimes the only
one amenable to explicit computation, like in Eq.~(\ref{eq:smoly})], and
the NLSM-like result~(\ref{eq:nlsm})
\begin{equation}\label{eq:nlsmfit}
  f_{\rm fit}^{\prime}(t)=C_p^{\prime} \exp[-\frac{1}{4\kappa_{\rm ELC}} \,
  \ln^3 (\kappa_{\rm ELC}t)].
\end{equation}
Using the fitted $\kappa$ from Fig.~\ref{fig:kappa}, we first
establish the boundaries of different intervals for $f(t)$ which
are discussed in Sec.~\ref{sec:qchaos}  (e.g., according to
Eq.~(\ref{eq:nlsm})  the beginning of the far tails is
approximately located at $t \gtrsim 1/\kappa$). These serves as a
consistent criterion, where portion of $f(t)$ to be fitted with
exp-log-cube formula analytical enlarges with increasing $W$,
thus avoiding spurious results when attempting to fit too
large piece of the distribution function. Although log-normal
asymptotics is expected in 2D systems~\cite{mirlin,smolyarenko}
(and has been confirmed numerically~\cite{uskiprb}) it has also
appeared as a candidate in 3D systems.~\cite{efetov} However,
attempt to fit $C_p^{\prime \prime} \exp(-C_2 \ln^2 t)$ to our
data fails completely, as shown in Figs.~\ref{fig:quest} and~\ref{fig:procent_elc}.

To check quantitatively the level at which fitted exp-log-cube formulas
match $f(t)$ from the numerical simulation, we perform our stringent
test by plotting the relative error in Fig.~\ref{fig:procent_elc}. The
possibility to fit both functions~(\ref{eq:nlsm}) and (\ref{eq:nlsmfit})
in a more or less similar way is shown in Figs.~\ref{fig:quest} and ~\ref{fig:procent_elc}.
Thus, a portion of $f(t)$ can be well-described by the exp-log-cube asymptotics, where
the size of the interval of amplitudes where the relative error is small (because of
the noise in the data points of $f(t)$ being pushed to the largest values of $t$)
increases with increasing disorder strength. Another conclusion which can be drawn from
Fig.~\ref{fig:procent_elc} is that a narrow intermediate  region of amplitudes exists which is not
covered by either the  FM function or exp-log-cube asymptotics. However, it
appears that it cannot be fitted at all  by the intermediate formula~(\ref{eq:interm}),
which is too close to  the PT distribution for this to work here.

\begin{figure}
\centerline{ \psfig{file=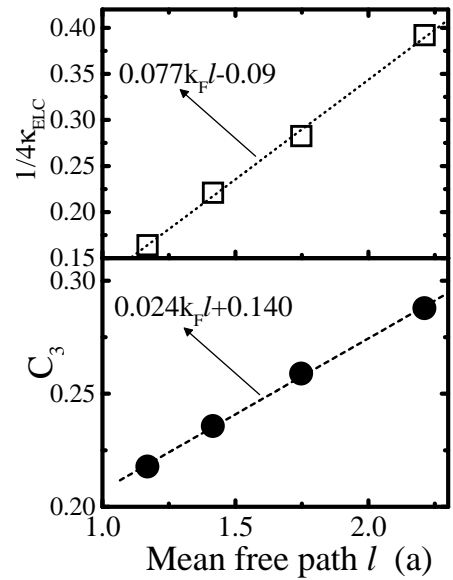,height=3.0in,angle=0} }
\vspace{0.2in} \caption{Parameters of the two different exp-log-cube
formulas [see Eqs.~(\ref{eq:elcfit}) and (\ref{eq:nlsmfit})],
fitted to the far tail of $f(t)$ in Fig.~\ref{fig:principal}, as a
function of disorder strength (measured by $\ell$). The success
of the linear fitting of both $C_3$ and $1/4\kappa_{\rm ELC}$ vs
$k_F\ell$ favors analytical predictions having first power of
$k_F\ell$ as the prefactor of the leading log-cube order, thereby
supporting DOF conclusions~\cite{smolyarenko} which emphasize that
short-scale effects govern the far tail in 3D systems.}
\label{fig:elcfits}
\end{figure}

We now proceed by analyzing dependence of the extracted
parameters on the disorder strength $k_F\ell$, as well as by
looking at the consequence of interpreting $\kappa_{\rm ELC}$ to
have the same physical meaning as $\kappa$ obtained by
fitting the FM formula to the perturbative region
(Fig.~\ref{fig:wl}). This interpretation is in the spirit of NLSM
conclusions where both perturbative and nonperturbative
corrections to RMT eigenlevel or eigenstate statistics are
governed by the same semiclassical physics.~\cite{mirlin} However,
their values turn out to be quite different along the respective
portions of $f(t)$. The most important finding in the far tail is
that both prefactors $C_3$ and $1/4\kappa_{\rm ELC}$  appear to increase
linearly with $k_F\ell$, as shown in Fig.~\ref{fig:elcfits}. Since prefactors
are the most conspicuous signature of the importance of
underlying ballistic or diffusive effects, this would mean that
ballistic effects completely dominate in the far tail (explaining
e.g., why we find $\kappa_{\rm ELC} \gg \kappa > \kappa_{\rm diff}$) and
confirming the conclusions of DOF analysis~\cite{smolyarenko}
where short-scale structure of the solutions of Schr\" odinger
equation is pointed out to be solely responsible for the rare
events at the largest possible intensities $t$. Nevertheless, a
puzzle remains: both prefactors depend on the sample size (which
is not treated explicitly in the present analytical
schemes~\cite{mirlin,smolyarenko}), whereas it is plausible that
ballistic effects are insensitive to $L$. For example, at $W=5$
we have to use $C_3=0.218$ or $\kappa_{\rm ELC}=1.411$ (compare
with corresponding values in Fig.~\ref{fig:elcfits}) to fit the far
tail of $f(t)$ in Fig.~\ref{fig:largecube} for the sample of size
$L=20a$. This suggests that a theory is needed where such
size-dependent effects are handled explicitly [with their
magnitude cannot be explained by, e.g., the change in
$\kappa_{\rm diff}(L)$ as is approximately possible for the size
dependence of parameters in the small-$t$ region].

\section{Conclusions}\label{sec:conclusion}

Our results confirm that statistical distribution of
eigenfunction amplitudes in 3D mesoscopic disordered systems
depends crucially on short-scale ballistic effects (and thereby
on the details of a random potential). In fact, the exactly
computed universal quantities characterizing the classical
diffusion process generate much smaller contribution to the
parameters needed to describe the observed distributions by the
analytically predicted formulas.
This is revealed in the
perturbative region of amplitudes, where deviations of the
eigenfunction amplitudes distribution function from the RMT limit
are small, and Fyodorov-Mirlin distribution (i.e., Porter-Thomas
distribution corrected by the weak localization terms) captures
their functional form for weak enough disorder
(i.e., the upper limit of disorder strength should be smaller than
the one determining the breakdown of other semiclassical
properties, like Drude-Boltzmann conductance) in the Anderson
model of localization. In this interval of small eigenfunction
amplitudes, the diffusive contribution contains boundary
conditions dependent weak localization terms, which stem from
evaluation of the discrete sum over the diffusion modes. In the
region of large wave function amplitudes, deviation from the RMT
appears in the form of far tail of the distributions function,
and is governed by the prelocalized states formed in rare
configurations of the random potential. The far tail, which we
obtain after examining about $2 \cdot 10^6$ exact eigenstates,
cannot be accounted by the log-normal distribution, but is
well-described by the exp-log-cube formulas. However,
substantially different parameters, used in NLSM formalism with
clear physical interpretation, are needed for this when compared
to the ones governing the small deviations. Nevertheless, their
linear dependence on the disorder strength (measured by
$k_F\ell$) is in accord with prefactor of the DOF theory, where
short-scale effects were pointed for the first time as the
possible sole explanation of the far tail statistics (we cannot
decipher whether these effects are of semiclassical or genuine
quantum origin). These findings appear to be quite different from
the success of various semiclassically based theories~\cite{uzy}
in describing the spectral statistics in mesoscopic systems, which
are dominated by the general properties of quantum coherent
superpositions and diffusion. The problems with standard diffusive
semiclassical description appear in the study of statistics of
any quantity where corrections to RMT are expressed in terms of
the sums which diverge above some specific space
dimensionality~\cite{basu} (such as $d \ge 2$ in the eigenstate
statistics problem). Our analysis points out that regularization
schemes, which {\em ad hoc} avoid divergences from the
short-length scales below $\ell$ by introducing a cutoff at large
momenta $|{\bf q}|\sim 1/\ell$, can lead to large discrepancies
when compared to exact numerical results.
\begin{figure}
\centerline{ \psfig{file=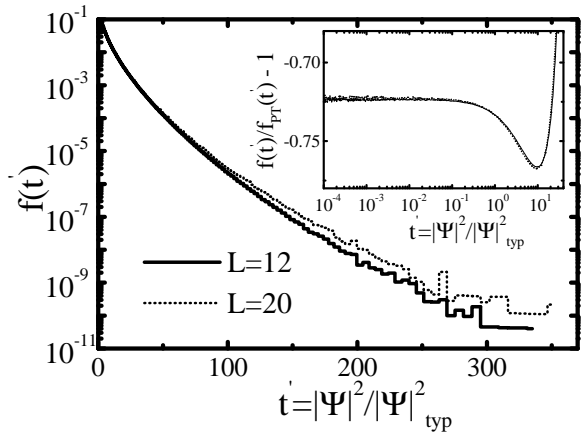,height=3.0in,angle=-90} }
\vspace{0.2in} \caption{Distribution function $f(t^\prime)$ of 
the eigenfunction intensities $t^\prime=|\Psi({\bf r})|^2/|\Psi({\bf r})|^2_{\rm typ}$ 
(normalized by the typical value of intensity $|\Psi({\bf r})|^2_{\rm typ}$ of 
a given eigenstate) in the Anderson modeled nanoscale conductors with $W=5$ disorder 
strength on $V=12^3$ and $V=20^3$ lattices. This analysis is complementary to that 
shown in Fig.~\ref{fig:largecube} for $f(t)$ where $t=|\Psi({\bf r})|^2/V^{-1}$. 
The inset plots correction [analogous to the one in Fig.~\ref{fig:wl} for 
$f(t)$] to the PT distribution $f_{\rm PT}(t^\prime)$ in the region of $t^\prime$ 
where such deviations are small.}
\label{fig:huse}
\end{figure}

The feature of the far tails in the Anderson model which does not
fit into the picture of ballistic effects alone is the need for
specific prefactors terms which can account for the observed size
dependence of the statistics. One possible explanation for these
discrepancies would be the lattice effects in our model, but real
solids are lattice structures (here simplified by taking single 
orbital per site) and possibility of such effects would not be an artifact 
of the model (which has been a paradigmatic model of the localization 
theory since the seminal paper of Anderson~\cite{anderson}). The other 
possibility stems from the fact that, even though our samples are good 
metals with $g > 10$, they are not large enough to be in the vicinity  
$g \gg 1$ of universality limit, which is the region treated by present 
theories. Strictly speaking, our findings can be considered as a demonstration 
of the structure of eigenfunctions in nanoscale conductors, and to 
contrast their structure to the predicted one in the limit $g \gg 1$, a 
heuristic attempt to account for the small-size corrections might be 
useful. Thus, if we interpret $t=|\Psi({\bf r})|^2 V$ of the standard statistical 
analysis as the ratio of eigenfunction intensity to its typical 
value $1/V$ far away from the high amplitude splashes, then by the same token, in 
small samples we can construct  statistics of analogous quantity  
$t^\prime=|\Psi({\bf r})|^2/|\Psi({\bf r})|^2_{\rm typ}$, where eigenfunction 
intensity is normalized by the its typical value $|\Psi({\bf r})|^2_{\rm typ}$. 
The auspicious outcome of this procedure would be vanishing of the size 
dependence of the far tail statistics. It turns out that, for the same  $W=5$ examples 
(on $12^3$ and $20^3$ cubic lattices) shown in Fig.~\ref{fig:largecube}, 
$|\Psi({\bf r})|^2_{\rm typ} V \approx 0.26$ fluctuates only slightly from eigenstate-to-eigenstate 
or from sample-to-sample. This allows us to define the PT distribution of $t^\prime$, 
$f_{\rm PT}(t^\prime)=[2 \pi t^\prime V |\Psi({\bf r})|^2_{\rm typ}]^{-1/2}\, \exp (-t^\prime 
V |\Psi({\bf r})|^2_{\rm typ}/2)$. The correction term $f(t^\prime)/f_{\rm PT}(t^\prime)-1$ 
in the region of small deviations from RMT now displays almost negligible size dependence 
(see inset in Fig.~\ref{fig:huse}). Nevertheless, Fig.~\ref{fig:huse} shows that size dependence 
of the far tail of $f(t^\prime)$ persists, albeit with a smaller relative difference 
$|f_{L=12}(t^\prime)-f_{L=20}(t^\prime)|/f_{L=12}(t^\prime)$ than in the case of usual 
analysis dealing with $f(t)$ in Fig.~\ref{fig:largecube}.

It remains to be seen if present approaches can combine short-scale 
effects with those responsible for the size dependence. Namely, ballistic extension of the
$\sigma$-model,~\cite{mirlinbal,balsigma} which attempts to
overcome the shortcomings of the diffusion approximation and
assumption of slow spatial variation of the NLSM fields, provides
a precise numerical coefficient of the log-cube prefactor while 
leaving its dependence on disorder strength unchanged from
$(k_F\ell)^2$ in the diffusive NLSM formalism---this contradicts
our findings. On the other hand, DOF technique as formulated in
Ref.~\onlinecite{smolyarenko} is not directly applicable to
the conductors modeled by the standard Anderson Hamiltonian.~\cite{smolyarenko}
Thus, the structure of eigenfunctions of the Schr\" odinger equation for
a particle in a random potential (which is the basic, tantalizingly
simply formulated one-particle quantum-mechanical problem) remains
a problem to be elucidated further.

Valuable discussions and suggestions during different stages of this 
project were provided by A. Bardas, V. Dobrosavljevi\' c, J. Fabian, 
A. D. Mirlin, and I. E. Smolyarenko. We are indebted to D. A. Huse for 
pointing out possible ways of grappling with small-size effects. 
B. N. acknowledges financial support from ONR grant N00014-99-1-0328.

$*$ Present address: Department of Physics and Astronomy, University of 
Delaware, Newark, DE 19716-2570.
%********************references***********************************


\begin{references}

\bibitem{meso} \emph{Mesoscopic phenomena in solids}, edited by B.L.
Altshuler, P.A. Lee, and R.A. Webb (North-Holland, Amsterdam,
1991).

\bibitem{mqp} \emph{Mesoscopic Quantum Physics}, edited by E. Akkermans,
J.-L. Pichard, and J. Zinn-Justin, Les Houches, Session LXI, 1994
(North-Holland, Amsterdam, 1995).

\bibitem{ucf} B.L. Al'tshuler, Pis'ma Zh. Eksp. Teor. Fiz. {\bf 41} 530 (1985)
[JEPT Lett. {\bf 41}, 648 (1985)]; P.A. Lee and A.D. Stone, Phys.
Rev. Lett. {\bf 55}, 1622 (1985).

\bibitem{shapiro} B. Shapiro, Phil. Mag. B {\bf 56}, 1031 (1987).

\bibitem{anderson} P.W. Anderson, Phys. Rev. {\bf 109}, 1492 (1958).

\bibitem{janssen} M. Janssen, Phys. Rep. {\bf  295}, 1 (1998).

\bibitem{mirlin} A.D. Mirlin,
Phys. Rep. {\bf 326}, 259 (2000).

\bibitem{fyodorov} Y.V. Fyodorov and A.D. Mirlin, Phys. Rev. B
{\bf 51}, 13 403 (1995).


\bibitem{mirlinsusy} A.D. Mirlin, J. Math. Phys. {\bf 38}, 1888
(1997).

\bibitem{efetov} V. I. Fal'ko and  K.B. Efetov, Europhys. Lett.
{\bf 32}, 627 (1995); V.I. Fal'ko and K.B. Efetov, Phys. Rev. B
{\bf 52}, 17 413 (1995).

\bibitem{smolyarenko} I.E. Smolyarenko and B.L. Altshuler, Phys. Rev.
B {\bf 55}, 10 451 (1997).

\bibitem{uskiprb} V. Uski, B. Mehlig, R.A. R\" omer, and M. Schreiber, Phys. Rev. B {\bf 62},
R7699 (2000).

\bibitem{bkn} B.K. Nikoli\' c, Phys. Rev. B {\bf 65}, 012201 (2002).

\bibitem{efetovbook} K.B. Efetov, \emph{Supersymmetry in Disorder and Chaos}
(Cambridge University Press, Cambridge, 1997).

\bibitem{mirlinbal}  Ya. M. Blanter, A. D. Mirlin, B. A.
Muzykantskii, Phys. Rev. B {\bf 63}, 235315 (2001).

\bibitem{smoly_private} I.E. Smolyarenko, private communication.

\bibitem{bouchaud} L. Laloux, P. Cizeau, J.-P. Bouchaud, and M.
Potters, Phys. Rev. Lett {\bf 83}, 1467 (1999); V. Plerou, P.
Gopikrishnan, B. Rosenow, L.A. Nunes Amaral, and H.E. Stanley,
Phys. Rev. Lett. {\bf 83} 1471 (1999).

\bibitem{lerner} B.L. Altshuler, V.E. Kravtsov, and I.V. Lerner,
in Ref.~\onlinecite{meso}.

\bibitem{muz} B.A. Muzykantskii and D.E. Khmelnitskii, Phys. Rev. B {\bf 51}, 5480 (1995).

\bibitem{basu} C. Basu, C.M. Canali, V.E. Kravtsov, and I.V.
Yurkevich, Phys. Rev. B {\bf 57}, 14 174 (1998).

\bibitem{sridhar} P. Pradhan and S. Sridhar, Phys. Rev. Lett. {\bf 85}, 2360 (2000).

\bibitem{allen} B.K. Nikoli\' c and P.B. Allen, Phys. Rev. B {\bf 63}, R020201 (2001).

\bibitem{dots} J.A. Folk \emph{et al.}, Phys. Rev. Lett. {\bf 76}, 1699 (1996);
A.M. Chang \emph{et al.}, Phys. Rev. Lett. {\bf 76}, 1695 (1996).

\bibitem{ogam} O. Agam, N.S. Wingreen, B.L. Altshuler, D.C. Ralph,
and M. Tinkham, Phys. Rev. Lett. {\bf 78}, 1956 (1997).

\bibitem{fmmodes} A. D. Mirlin and Y. V. Fyodorov, J. Phys. A {\bf 26}, L551 (1993);
Y. V. Fyodorov and A. D. Mirlin, Int. J. Mod. Phys. B {\bf 8},
3795 (1994).

\bibitem{ensemble} Note that in this paper the word {\bf ensemble}
denotes our numerically generated set of conductors with different
impurity configurations but of the same disorder strength (i.e.,
impurity concentration). This should not be confused with the word
{\em ensemble} which denotes in random matrix theory a set of all
random matrices obeying specific symmetries---the time-reversal
and spin-rotation invariant Hamiltonians from each of our ``five
ensembles'' belong to the Gaussian orthogonal {\em ensemble} of
RMT.



\bibitem{chaos} \emph{Chaos in Quantum Physics}, edited by M.-J. Jianonni, A.
Voros, and J. Zinn-Justin, Les-Houches, Session LII, 1989
(North-Holland, Amsterdam, 1991).

\bibitem{ghur} T. Ghur, A. M\" uller-Groeling,
and H.A. Widenm\" uller, Phys. Rep. {\bf 299}, 189 (1998).

\bibitem{bohigas} O. Bohigas, M. J. Giannoni, and C. Schmit, Phys.
Rev. Lett. {\bf 52}, 1 (1984).

\bibitem{andreev} A.V. Andreev, O. Agam, B.D. Simons, and B.L.
Altshuler, Phys. Rev. Lett. {\bf 76}, 3947 (1996).

\bibitem{gorkov} L. P. Gor'kov and G. M. Eliashberg, Zh. Eksp. Teor. Fiz. {\bf
48}, 1407 1965 [Sov. Phys. JETP {\bf 21} 1965].


\bibitem{shklovskii} B. L. Altshuler and B. I. Shklovskii, Zh. Eksp. Teor. Fiz.
{\bf 91}, 220 (1986) [Sov. Phys. JETP {\bf 64}, 127 (1986)].

\bibitem{smith} R.S. Whitney, I.V. Lerner, R.A. Smith, Waves in Random Media
{\bf 9}, 179 (1999).


\bibitem{prange} R. E. Prange, Phys. Rev. Lett. {\bf 78}, 2280
(1997).


\bibitem{simons} A. Altland, C. R. Offer, and B. D. Simons, in
{\em Supersymmetry and Trace Formula}, ed. by I. V. Lerner, J. P.
Keating, and D. E. Khmelnitskii, NATO ASI series B, Vol. 370
(Kluwer Publishing, Dodrecht, 1999).

\bibitem{fyodorov_rmt} Y. V. Fyodorov, in Ref.~\onlinecite{mqp}.

\bibitem{porter} C. E. Porter and R. G. Thomas, Phys. Rev. {\bf 104}, 483 (1956).

\bibitem{haake} F. Haake, {\em Quantum Signatures of Chaos}
(Springer-Verlag, Heidelberg, 2001).


\bibitem{kaplan} L. Kaplan, Phys. Rev. Lett. {\bf 80}, 2582 (1998).

\bibitem{wl} L. P. Gor'kov, A. I. Larkin, D. E. Khmel'nitskii, Pis'ma Zh. Eksp.
Teor. Fiz. {\bf 30}, 248 (1979) [JETP Lett. {\bf 30}, 228 (1979)].

\bibitem{montsum} D. Braun, E. Hofstetter, G. Montambaux, A.
MacKinnon, Phys. Rev. B {\bf 64}, 155107 (2001).

\bibitem{balsigma} B. A. Muzykantskii and D. E. Khmelnitskii,
Pisma Zh. Eksp. Teor. Fiz. {\bf 62}, 68 (1995) [JETP Lett. {\bf
62}, 76 (1995)].

\bibitem{uski2001} V. Uski, B. Mehlig, and M. Schreiber, Phys. Rev. B {\bf 63}, 241101 (2001).

\bibitem{uski} V. Uski, B. Mehlig, R.A. R\"omer, M. Schreiber, Ann. Phys.
(Leipzig) {\bf 7}, 437 (1998).

\bibitem{monty} G. Montambaux, cond-mat/0012210.

\bibitem{mirlinprb} A.D. Mirlin, Phys. Rev. B {\bf 53}, 1186
(1996).

\bibitem{dek} B.A. Muzykantskii and D.E. Khmelnitiskii, cond-mat/9601045.

\bibitem{lb} R. Landauer, Phil. Mag. {\bf
21}, 863 (1970); C. Caroli, R. Combescot, P. Nozieres, and D.
Saint-James, J. Phys C {\bf 4}, 916 (1971).

\bibitem{datta} S. Datta,
\emph{Electronic Transport in Mesoscopic Systems} (Cambridge
University Press, Cambridge, 1995).

\bibitem{mackinnon} D. Braun, E. Hofstetter, A. MacKinnon, and G. Montambaux,
Phys. Rev. B {\bf 55}, 7557 (1997).

\bibitem{nikolic_qpc} B. K. Nikoli\' c and P. B. Allen, J. Phys.: Condens.
Matter {\bf 12}, 9629 (2000).

\bibitem{sharvin} Yu.V. Sharvin, Zh. Eksp. Teor. Fiz. {\bf 48}, 984 (1965)
[Sov. Phys. JETP {\bf 21}, 655 (1965)].


\bibitem{kubo} B.K. Nikoli\' c, Phys. Rev. B {\bf 64}, 165303 (2001).


\bibitem{uzy} N. Argaman, Y. Imry, and U. Smilansky, Phys. Rev. B {\bf 47},
4440 (1993).

\end{references}
\end{document}